\documentstyle[prl,aps,epsfig,floats]{revtex}

\begin{document}
\draft

\twocolumn[\hsize\textwidth\columnwidth\hsize\csname @twocolumnfalse\endcsname
\title{ \bf Diagrammatic Quantum Monte Carlo for Two-Body
Problem: Exciton}
\author{E.\ A.\ Burovski$^{1}$,
A.\ S.\ Mishchenko$^{2,1}$, N.\ V.\ Prokof'ev$^{3}$,
and B.\ V.\ Svistunov$^{1}$}
\address{
$^1$RRC 'Kurchatov Institute', 123182, Moscow, Russia \\
$^2$Correlated Electron Research Center, Tsukuba Central 4,
Tsukuba 305-8562, Japan \\
$^3$Department of Physics, University of
Massachusets, Amherst, Masachusets 01003}

\maketitle
\begin{abstract}
We present a novel method for precise numerical solution of the
irreducible two-body problem and apply it to excitons in solids.
The approach is based on the Monte Carlo simulation of the two-body
Green function specified by Feynman's diagrammatic expansion. Our
method does not rely on the specific form of the electron and hole
dispersion laws and is valid for any attractive electron-hole
potential. We establish limits of validity of the Wannier (large
radius) and Frenkel (small radius) approximations, present
accurate data for the intermediate radius excitons, and give
evidence for the charge transfer nature of the monopolar exciton
in mixed valence materials.
\end{abstract}
\vskip1pc] \narrowtext

After it was realized that under certain conditions the electron dynamics in
conduction band is of two-particle nature due to Coulomb attaction to the
hole in the valence band left behind \cite{Fr31}, the problem of exciton
became a model example of an irreducible (center-of-mass motion does not
separate from the rest of degrees of freedom) two-body problem. The simplest
(still rather general) exciton Hamiltonian \cite{Egri85,Knox} consists of
conduction and valence band contributions, $H_{0}$, and coupling $H_{%
\mbox{\scriptsize e-h}}$:
\begin{equation}
H_{0}=\sum_{{\scriptsize {\bf k}}}\varepsilon _{c}({\bf k})e_{{\scriptsize
{\bf k}}}^{\dagger }e_{{\scriptsize {\bf k}}}+\sum_{{\bf k}}\varepsilon _{v}(%
{\bf k})h_{{\scriptsize {\bf k}}}h_{{\scriptsize {\bf k}}}^{\dagger },
\label{2}
\end{equation}
\begin{equation}
H_{\mbox{\scriptsize e-h}}=-N^{-1}\sum_{{\scriptsize {\bf pkk^{\prime }}}}%
{\cal U}({\bf p},{\bf k},{\bf k}^{\prime })e_{{\scriptsize {\bf p+k}}%
}^{\dagger }h_{{\scriptsize {\bf p-k}}}^{\dagger }h_{{\scriptsize {\bf %
p-k^{\prime }}}}e_{{\scriptsize {\bf p+k^{\prime }}}}.  \label{3}
\end{equation}
Here $e_{{\bf k}}$ ($h_{{\bf k}}$) is the electron (hole) annihilation
operator, $\varepsilon _{c}({\bf k})$ ($\varepsilon _{v}({\bf k})$) is the
conduction (valence) band dispersion law, N is the number of lattice sites,
and ${\cal U}({\bf p},{\bf k},{\bf k}^{\prime })$ is an attractive
interaction potential.

Despite numerous efforts over the years there is no rigorous technique to
solve for exciton properties even for the simplest model given above which
treats electron-electron interactions as a static renormalized Coulomb
potential with averaged dynamical screening. The only solvable cases are the
Frenkel small-radius limit \cite{Fr31} and the Wannier large-radius limit
\cite{Wan37} which describe molecular crystals and wide gap insulators with
large dielectric constant, respectively. Much more frequently encountered
cases of intermediate radius excitons (e.g. intermediate gap semiconductors,
LiF, or mixed valence systems) have to be dealt with using approximate
numerical approaches. There are powerful {\it ab initio} modern methods \cite
{Alb98,Ben98,RohLo98} for band structure and effective electron-hole
potential calculations, but the real bottleneck is in numerical solution of
the two-particle problem for a bulk material. One can either solve the
Bethe-Salpeter equation on a finite mesh in reciprocal/direct space \cite
{Alb98,Ben98,RohLo98}, or employ the random-phase approximation decoupling
\cite{Egri85}. However, both methods suffer from systematic errors, and the
Bethe-Salpeter equation on finite mesh may lead to incorrect eigenstates 
for the Wannier case \cite{RohLo98}. Therefore, even the
limits of validity of the Wannier and Frenkel approximations can not be
established by existing methods.

Besides, an efficient and rigorous method for the study of exciton properties,
given the band structure, is of high virtue for phenomenological models. As
an example, we refer to the protracted discussion of numerous (and often
contradictory) models concerning exciton properties in mixed valence
semiconductors \cite{CuKiMi}. In Ref.~\onlinecite{KiMi90} unusual properties
of SmS and SmB$_6$ were explained by invoking the excitonic instability
mechanism assuming charge-transfer nature of the optically forbidden
exciton. Although this model explains quantitatively the phonon spectra \cite
{KiMi91}, optical properties \cite{TraWa84,Le95}, and magnetic neutron
scattering data \cite{KiMi95}, its basic assumption has been criticized as
being groundless \cite{Kas94}.

In this Letter we describe how ground state properties of excitons in the
model (\ref{2})-(\ref{3}) can be obtained numerically without systematic
errors for arbitrary dispersion relations $\varepsilon _{c}({\bf k})$ and $%
\varepsilon _{v}({\bf k})$), and attractive potential ${\cal U}({\bf p},{\bf %
k},{\bf k}^{\prime })$. First, we show that the problem fits into the
diagrammatic Monte Carlo (MC) method \cite{PS98,MPSS00,QDST01} which sums
positively-definite perturbation series, in our case Feynman diagrams, for
the two-particle Matsubara Green function, $G$. We then describe the
procedure of extracting various physical properties from the asymptotic
long-time behavior of $G$. Next, we discuss our results for a particular
tight-binding model and electron-hole interaction potential to see under
what conditions Frenkel and Wannier approximations remain accurate. Finally,
we present evidence that the band structure of mixed valence materials results
in the charge-transfer character of the optically forbidden exciton.

The two-particle Green function with total momentum $2{\bf p}$ in imaginary
time representation is defined as
\begin{equation}
G_{{\scriptsize {\bf p}}}^{{\scriptsize {\bf kk}}^{\prime }}(\tau )=\langle
0\mid e_{{\scriptsize {\bf p+k^{\prime }}}}(\tau )h_{{\scriptsize {\bf %
p-k^{\prime }}}}(\tau )h_{{\scriptsize {\bf p-k}}}^{\dagger }e_{{\scriptsize
{\bf p+k}}}^{\dagger }\mid 0\rangle ,  \label{4}
\end{equation}
where the vacuum state $\mid 0\rangle $ corresponds to empty conduction and
filled valence bands, and $h_{{\bf p-k}}(\tau )=e^{H\tau }h_{{\bf p-k}%
}e^{-H\tau }$, $\tau >0$. In the interaction representation $G$ can be
written as a sum of ladder-type Feynman diagrams, see Fig.~\ref{fig:fig1}:
pairs of horizontal solid lines represent free electron-hole pair
propagators, $G_{{\bf p}}^{(0)}({\bf k},\tau _{2}-\tau _{1})=\exp \left(
-\varepsilon ({\bf k})(\tau _{2}-\tau _{1})\right) $, where $\varepsilon (%
{\bf k})=\varepsilon _{c}({\bf p+k})-\varepsilon _{v}({\bf p-k})$ is the
energy of the pair, and dashed lines represent the interaction potential.
\begin{figure}[th]
\begin{picture}(100,90)
\put(0,5){\line(1,0){240}}
\put(0,70){\line(1,0){240}}
\multiput(40,70)(0,-10){7}{\line(0,-1){5}}
\multiput(110,70)(0,-10){7}{\line(0,-1){5}}
\multiput(180,70)(0,-10){7}{\line(0,-1){5}}
\put(0,75){$0$}
\put(40,75){$\tau_1$}
\put(110,75){$\tau_2$}
\put(180,75){$\tau_3$}
\put(235,75){$\tau$}
\put(5,55){${\bf p}+{\bf k}$}
\put(5,13){${\bf p}-{\bf k}$}
\put(60,55){${\bf p}+{\bf k}_1$}
\put(60,13){${\bf p}-{\bf k}_1$}
\put(130,55){${\bf p}+{\bf k}_2$}
\put(130,13){${\bf p}-{\bf k}_2$}
\put(198,55){${\bf p}+{\bf k}'$}
\put(198,13){${\bf p}-{\bf k}'$}
\put(45,31){$V_1(2{\bf p})$}
\put(115,31){$V_2({\bf k}_2-{\bf k}_1)$}
\put(185,31){$V_2({\bf k}'-{\bf k_2})$}
\end{picture}
\caption{A typical diagram contributing to $G_{{\bf p}}^{{\bf kk}^{\prime }}(%
\protect\tau )$.}
\label{fig:fig1}
\end{figure}
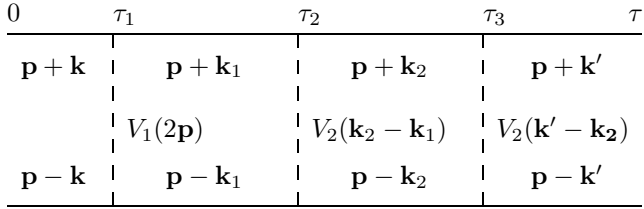
For purely numerical reasons explained below we split the potential into two
terms ${\cal U}({\bf p},{\bf k},{\bf k}^{\prime })=V_{1}(2{\bf p})+V_{2}(%
{\bf k}-{\bf k}_{1})$, and expand in both $V_{1}$ and $V_{2}$. [This can be
done because \cite{Egri85} ${\cal U}({\bf p},{\bf k},{\bf k}^{\prime
})=V_{0}-W(2{\bf p})+U({\bf k-k^{\prime }})$ where $V_{0}$ is the on-site
coupling, $W(2{\bf p})$ is the dipolar term, $U({\bf k-k^{\prime }}%
)=\sum_{\lambda \neq 0}\exp (i{\bf qR}_{\lambda })/(R_{\lambda }\epsilon (%
{\bf R}_{\lambda }))$ is the monopolar term, and $\epsilon ({\bf R}_{\lambda
})$ is a static dielectric screening function (we set the electric charge to
unity). Since $U({\bf q})$ is not positive definite [in fact $\sum_{{\bf q}%
}U({\bf q})=0$] we add and subtract some constant ${\bar{U}}$ to ensure that
$V_{1}(2{\bf p})=V_{0}-W(2{\bf p})-{\bar{U}}$ and $V_{2}({\bf q})={\bar{U}}%
+U({\bf q})$ are both positive - this imposes the only limitation on value $%
V_{0}$ in our method].

The final answer for $G$ is given by the sum of all possible diagrams.
Formally we can write this as a series of multi-dimensional integrals
\[
G_{{\scriptsize {\bf p}}}^{{\scriptsize {\bf kk}}^{\prime }}(\tau
)=\sum_{m=0}^{\infty }\sum_{\xi _{m}}\int dx_{1}\cdots dx_{m}\,F_{%
{\scriptsize {\bf p}}}^{{\scriptsize {\bf kk}}^{\prime }}(\tau ;\xi
_{m};x_{1},\ldots ,x_{m}).
\]
where $x_{1},\dots x_{m}$ are internal variables [times and momenta, $%
x_{i}=(\tau _{i},{\bf k}_{i})$] of the $m$-th order diagram, the summation
over $\xi _{m}$ accounts for different diagrams of order $m$, and the
``weight'' $F$ is given by the product of electron-hole propagators and
interaction vertices according to standard rules. For positive $V_{1}$ and $%
V_{2}$ all terms in the series are positive definite and one may apply the
diagrammatic Monte Carlo technique developed in Refs.~%
\onlinecite{PS98,MPSS00}, which evaluates such series without systematic
errors (by Metropolis-type sampling of diagrams according to their weight
directly in the momentum-time continuum). Since the method itself is well
described in the literature we will concentrate on the problem specific
details only.

The crucial for the whole scheme update is the one which changes the number
of interaction vertices by one. To render algorithm efficient one has to
propose new internal parameters as close as possible to the distribution
function $R(x_{m+1}) = F(\xi_{m+1} ; x_1, \ldots , x_m,x_{m+1})/ F(\xi_{m} ;
x_1, \ldots , x_m)$ defined by the ratio of the new and old diagram weights.
This is done in order to maximize the acceptance ratio $P_{{\rm acc}}$ ---
if proposed $x_{m+1}=(\tau_{m+1}, {\bf k}_{m+1})$ are distributed according
to some normalized function $W(x_{m+1})$, then $P_{{\rm acc}} \propto
R(x_{m+1})/W(x_{m+1}) $. Otherwise the choice of $W(x_{m+1})$ is a matter of
computational convenience \cite{PS98,MPSS00}.

First, we select (with equal probabilities) which interaction vertex, $V_{1}$
or $V_{2}$, will be inserted, and then select at random the time interval
where it will be placed, $\tau _{m+1}\in (\tau _{a},\tau _{b})$, where $\tau
_{a,b}$ are the interval boundaries determined either by the existing
interaction vertices or the diagram ends. All the momenta at $\tau <\tau _{a}
$ and $\tau >\tau _{m+1}$ are kept untouched. In case when $V_{1}(2{\bf p})$
is inserted, the new momentum ${\bf k}_{m+1}$ is proposed uniformly in the
Brillouin zone (BZ). When $V_{2}({\bf k}_{b}-{\bf k}_{m+1})$ is inserted ($%
{\bf k}_{b}$ is the relative motion momentum to the left of point $(\tau _{b}
$) the new momentum ${\bf k}_{m+1}$ is proposed using distribution function $%
W({\bf k}_{m+1})=\left( \beta /2\pi \arctan \beta \right) ^{3}\prod_{\alpha
}\,\left( 1+\beta k_{m+1}^{(\alpha )}/\pi )^{2}\right) ^{-1}$, where $\alpha
=x,y,z$. The parameter $\beta $ is uniformly seeded on interval $[\beta _{%
{\rm min}},\beta _{{\rm max}}]$ at each step, and $\beta _{{\rm min}},\beta
_{{\rm max}}$ are further tuned to maximize the acceptance ratio. We note,
that different distribution functions used to propose new momentum ${\bf k}%
_{m+1}$ when dealing with $V_{1}$ and $V_{2}$ vertices was the only
motivation behind an artificial separation ${\cal U}=V_{1}+V_{2}$ [The
actual gain in efficiency was about three orders of magnitude!]. Finally,
the time position for the new vertex was seeded according to the
distribution function dictated by the diagram weights ratio $W(\tau
_{m+1})=\delta \epsilon \cdot e^{-\delta \epsilon \tau _{m+1}}/(e^{-\delta
\epsilon \tau _{a}}-e^{-\delta \epsilon \tau _{b}})$ where, $\delta \epsilon
=\varepsilon _{{}}({\bf k}_{m+1})-\varepsilon _{{}}({\bf k}_{b})$.

We also employ standard Metropolis updates changing the values of internal
momenta and times,
which substantially enhances the efficiency of the
algorithm.

\begin{figure}[ht]
\epsfxsize=0.47\textwidth \epsfbox{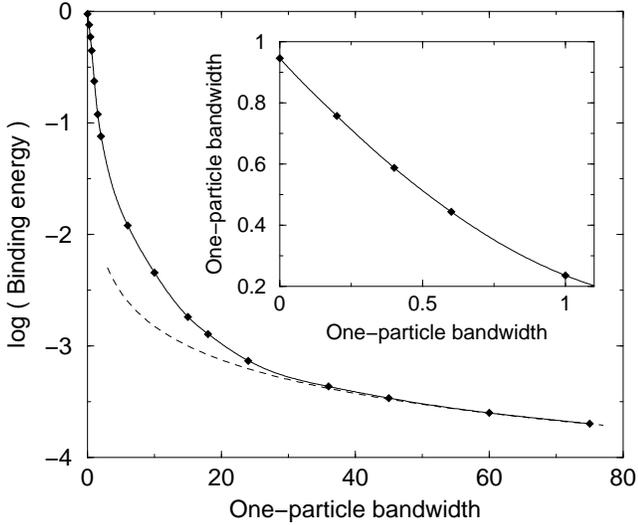}
\caption{The dependence of the exciton binding energy on the bandwidth $E_c=E_v$.
 Statistical errors are less than $5 \cdot 10^{-3}$ in relative
units. The dashed line corresponds to the Wannier model.  The solid line is
the cubic spline, the derivatives at the right and
left ends being fixed by the Wannier limit and perturbation
theory, respectively. Insert: the initial part of the plot.}
\label{figr1}
\end{figure}
We now turn to the discussion of how exciton properties are
obtained from the $G (\tau \to \infty )$ limit. An eigenstate
$\mid \nu; {\bf p} \rangle$ with energy $E_{\nu}$ can be written
as
\begin{equation}
\mid \nu; {\bf p } \rangle \equiv \sum_{{\scriptsize {\bf k}}} \xi_{%
{\scriptsize {\bf p k}}}(\nu) e^{\dagger}_{{\scriptsize {\bf p+k}}}
h^{\dagger}_{{\scriptsize {\bf p-k}}} \mid 0 \rangle.  \label{WF}
\end{equation}
where amplitudes $\xi_{{\bf p k}}(\nu)= \langle \nu; {\bf p} \mid
e^{\dagger}_{{\bf p+k}} h^{\dagger}_{{\bf p-k}}  \mid 0 \rangle$
describe the wave function of internal motion of the exciton. In
terms of exciton eigenstates we have, $G_{{\bf p}}^{{\bf
k=k}^{\prime}} (\tau) = \sum_{\nu} \mid \xi_{{\bf p k}}(\nu)
\mid^2 e^{-E_{\nu} \tau}$, and if $\tau $ is much larger than
inverse energy difference between the ground and first excited
states, the Green function projects to the ground state, $G_{{\bf p}}^{{\bf %
k=k}^{\prime}} (\tau \to \infty ) = \mid \xi_{{\bf p k}}(\mbox{g.s.}) \mid^2
e^{-E_{\mbox{\scriptsize g.s.}} \tau }$. Due to normalization condition $%
\sum_{{\bf k}} \mid \xi_{{\bf p k}}(\nu ) \mid^2 \equiv 1$ the asymptotic
behavior of the sum $\tilde{G}_{{\bf p}} = \sum_{{\bf k}} G_{{\bf p}}^{{\bf %
k=k}^{\prime}} $ is especially simple, $\tilde{G}( \tau ) \to e^{-E_{%
\mbox{\scriptsize g.s.}} \tau}$. This asymptotic behavior allows simulations
of energy and amplitudes at fixed $\tau$ [large enough to make the 
corresponding systematic error negligible], using the technique of Monte Carlo
estimators. To this end we differentiate each diagram for $\tilde{G}( \tau )$
\cite{rem:1} with respect to $\tau $ and arrive at the result (compare with Ref.~\cite{MPSS00})
\begin{equation}
E_{\mbox{\scriptsize g.s.}} = \tau^{-1} \left\langle \sum_{j=1}^{m+1}
\varepsilon^{j}({\bf k}) \Delta\tau_j \; - \; m \right\rangle_{%
\mbox{\scriptsize MC}} \;,  \label{estene}
\end{equation}
where $\langle ... \rangle_{\mbox{\scriptsize MC}}$ stands for the MC
statistical average, $m$ is the diagram order, $\varepsilon^{j}({\bf k})$
and $\Delta\tau_j$ are the electron-hole pair energy and duration of the $j$%
-th propagator, respectively. By definition, in the limit $\tau \to \infty$
we have $G_{{\bf p}}^{{\bf k=k}^{\prime}} /\tilde{G}_{{\bf p}} = \mid \xi_{%
{\bf p k}}(\mbox{g.s.}) \mid^2$, i.e. the distribution over quasimomentum $%
{\bf k}$ is related to the wave function of internal motion. The wave
function of the bound state can be chosen real, and the Fourier transform
may be used to obtain $\mid \mbox{g.s.} \rangle$ in direct space \cite{rem:0}%
. 
\begin{figure}[hbt]
\epsfxsize=0.47\textwidth \epsfbox{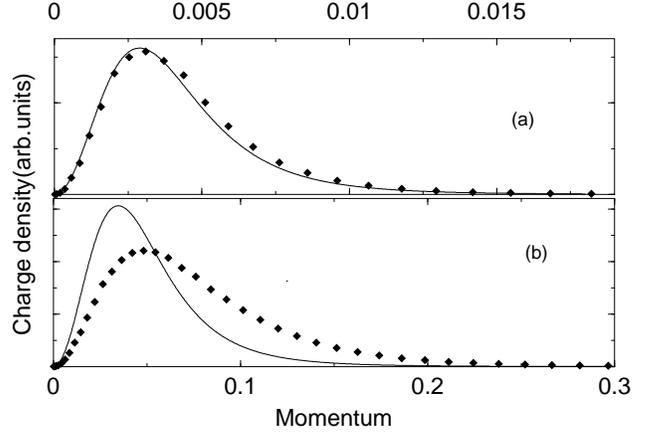}
\caption{The momentum dependence of the charge density $\mid \protect\xi_{%
{\bf p k}}(\mbox{g.s.}) \mid^2 k^2$ for $E_c=E_v=60$ (a) and $E_c=E_v=10$
(b). Solid lines are the Wannier model result. Statistical errors are
typically of order $10^{-4}$. }
\label{figr3}
\end{figure}

In this Letter we focus on the study of exciton properties in a
simple cubic 3D lattice with tight binding dispersion laws for the
electron and hole bands
\begin{equation}
\varepsilon _{c,v}({\bf k})=\tilde{E}_{c,v}\pm (E_{c,v}/6)\sum_{\alpha
}(1-\cos k_{\alpha }).  \label{Eeh}
\end{equation}
The choice of interaction parameters was motivated by the possibility to
cover all regimes (from Wannier to Frenkel limit) by varying the ratio
between the bandwidth and the gap only. Our simulations were done for $%
\tilde{E}_{v}=0$, $E_{g}\equiv \tilde{E}_{c}=1$, $W(2{\bf p}=0)=-0.168$, $%
V_{0}=0.778$, ${\bar{U}}=0.578$, and $\epsilon ({\bf R})=10$ \cite{Evald}.
The binding energy in the Frenkel limit $E_{\mbox{\scriptsize FL}}$ ($%
E_{c,v}\ll E_{g}$) is then less than the gap, $E_{\mbox{\scriptsize FL}%
}=V_{1}(2{\bf p}=0)+\sum_{{\bf q}}V_{2}({\bf q})=0.946$, thus rendering the
exciton stability for all values of $E_{c,v}$. In the Wannier limit of large
bandwidth $E_{c,v}\gg E_{g}$ the binding energy approaches $3/(2\epsilon
^{2}E_{c})$ (assuming $E_{v}=E_{c}$). Of course, our parameters satisfy the
requirement that $V_{1}$ and $V_{2}$ are positive definite functions.

\begin{figure}[ht]
\epsfxsize=0.47\textwidth 
\epsfbox{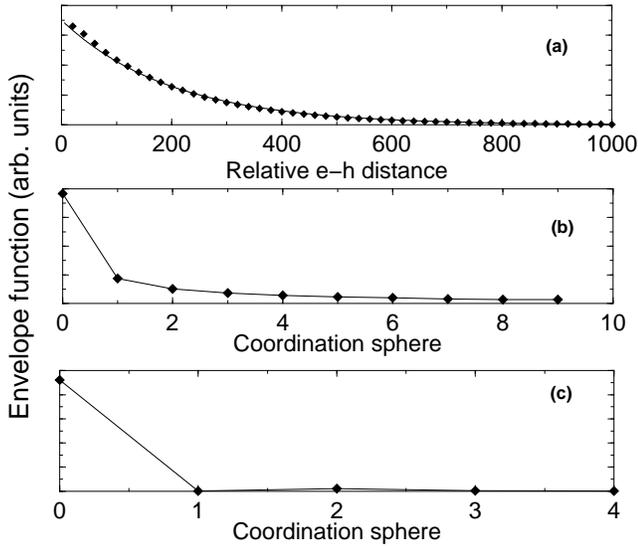}
\caption{ The wave function of internal motion in real space: (a) Wannier 
[$E_c=E_v=60$]; (b) intermediate [$E_c=E_v=10$]; (c) near-Frenkel
[$E_c=E_v=0.4 $] regimes. 
The solid line in the panel
(a) is the Wannier model result while solid lines in other panels
are to guide an eye only. Statistical errorbars are of order
$10^{-4}$. 
} 
\label{figr4}
\end{figure}

Our results for the binding energy and wave function are shown in Figs. ~\ref
{figr1}, ~\ref{figr3}, and \ref{figr4}. First we notice that the method
works equally well in all regimes, and statistical errors are much smaller
than symbols sizes in all plots. An unexpected result is that extremely
large bandwidth $E_c/E_g >20$ is necessary for the Wannier approximation to
be adequate: both the binding energy $E_B$ (Fig.\ \ref{figr1}) and the wave
function \cite{rem:3} 
(Fig.\ \ref{figr3} and Fig.\ \ref{figr4} (b)) demonstrate large
deviations for smaller $E_c/E_g$. Most surprisingly, for $1 < E_c/E_g < 10$
the wave function has a large (and dominating) on-site component [Fig.\ \ref
{figr4}(b)], but the binding energy is not even close to the Frenkel limit!
For $E_c/E_g=0.4$ the wave function is almost entirely localized [Fig.\ \ref
{figr4}(c)] but $E_{\mbox{g.s.}}$ is still 50\% away from the small-radius
limit. Noticing that $E_{\rm g.s.} \approx E_{\mbox{\scriptsize FL}}-(E_c+E_v)/2$ ($%
E_c=E_v$), which holds for localised functions when $E_c<0.4$, we deduce 
that the deviation from Frenkel result is
determined  by the electron and hole delocalization energy. Our
conclusion is then that the intermediate-range regime is very broad and
relevant in most practical cases. 
\begin{figure}[ht]
\epsfxsize=0.47\textwidth \epsfbox{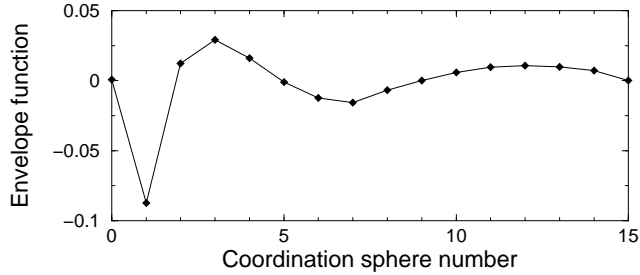} \caption{The wave
function of internal motion in real space for the optically
forbidden monopolar ($W(2{\bf p})=0$) exciton defined by the
following model parameters: $\tilde{E}_c=1.5$, $\tilde{E}_v=0$, $E_c=-0.5$, $%
E_v=0.05$, $\protect\epsilon=10$, $V_0=0.578$. Statistical errorbars are of
order $10^{-4}$. }
\label{figr5}
\end{figure}

To study the structure of optically forbidden excitons in mixed valence
compounds we choose typical for these semiconducting materials band spectra
\cite{CuKiMi}, i.e. an almost flat valence band separated by an indirect gap
from the wide conduction band with maximum at ${\bf k}=0$ and minimum at the
BZ boundary. One can see in Fig.~\ref{figr5} that this leads to the charge
transfer character of the optically forbidden monopolar exciton ($W(2{\bf p}%
)=0$) when the wave function of internal motion has 
almost zero on-site component, maximal charge density at near neighbours,
and large long-ranged oscillations at neighboring sites. The difference with
the previously discussed $E_{v,c}/E_g=0.4$ case, see Fig.~\ref{figr4}(b), 
is remarkable.

Finally, we would like to note that diagrammatic MC technique not
only gives properties of the ground state but is also suitable for
the study of excited states and optical absorption \cite{rem:4}.
This can be done by simulating the $\tau $
dependence of ${\cal G}({\bf p=0}, \tau )= \sum_{{\bf kk}^{\prime}} G_{{\bf p%
}=0}^{{\bf kk}^{\prime}} (\tau )$, and solving numerically equation
\[
{\cal G}({\bf p=0}, \tau)= \int_0^{\infty} g(\omega) \exp(-\omega \tau) d
\omega
\]
to obtain the spectral function $g(\omega)$ \cite{MPSS00}.


We thank A.\ Sakamoto and N.\ Nagaosa for fruitful discussions. We
acknowledge the support of the National Science Foundation under Grant
DMR-0071767 and RFBR grant 01-02-16508.


\begin{references}
\bibitem{rem:0}  If the quantity $\mid \xi_{{\bf p k}}(\mbox{g.s.}) \mid^2$
has no nodes, the amplitudes $\xi_{{\bf p k}}(\mbox{g.s.})$ can be chosen
positive. Since $G^{{\bf k}{\bf k}'}_{\bf p} (\tau ) $ is given by the 
sum of positive-definite diagrams we know that 
$\xi_{{\bf p k}}(\mbox{g.s.})$ has no nodes.
\bibitem{rem:3}  While for Wannier regime the wave function is spherically
symmetric, it's symmetry for the Frenkel and intermediate case regimes bears
the lattice structure thus making the coordination spheres number a natural
unit for the representation of the wave function in direct space.
\bibitem{rem:4} In the case when the interband optical matrix element is momentum
independent, the absorption coefficient is proportional
to $g(\omega)$.
\end{references}
\end{document}